\documentclass[lnicst]{svmultln}

\usepackage{makeidx}  
\usepackage[utf8]{inputenc}
\usepackage{cite}
\usepackage{subfigure}
\usepackage{graphicx}
\usepackage{epstopdf}
\begin{document}
\mainmatter              
\title{Modeling Epidemic Risk Perception in Networks with Community Structure}
\titlerunning{Modeling Risk Perception in Networks with Community Structure}  
%
\author{Franco Bagnoli\inst{1,2,3} \and Daniel Borkmann\inst{4} \and
Andrea Guazzini\inst{5,6} \and \newline Emanuele Massaro\inst{7} \and
Stefan Rudolph\inst{8}}
\authorrunning{Franco Bagnoli et al.}   
%
\tocauthor{Franco Bagnoli, Daniel Borkmann, Andrea Guazzini, Emanuele Massaro, Stefan Rudolph}
\institute{Department of Energy, University of Florence, Italy\\
\and Center for the Study of Complex Systems, University of Florence, Italy\\
\and National Institute for Nuclear Physics, Florence Section, Italy\\
\and Communication Systems Group, ETH Zurich, Switzerland\\
\and National Research Council, Institute for Informatics and Telematics, Pisa, Italy\\
\and Department of Psychology, University of Florence, Italy\\
\and Department of Computer Science and Systems, University of Florence, Italy\\
\and Organic Computing Group, University of Augsburg, Germany
}

\toctitle{Lecture Notes in Computer Science}
\tocauthor{Franco Bagnoli et al.}
\maketitle              
\begin{abstract}        
We study the influence of global, local and community-level risk perception on
the extinction probability of a disease in several models of social networks.
In particular, we study the infection progression as a susceptible-infected-susceptible
(SIS) model on several modular networks, formed by a certain number of random and
scale-free communities. We find that in the scale-free networks the progression is
faster than in random ones with the same average connectivity degree. For what
concerns the role of perception, we find that the knowledge of the infection level
in one's own neighborhood is the most effective property in stopping the spreading
of a disease, but at the same time the more expensive one in terms of the quantity
of required information, thus the cost/effectiveness optimum is a tradeoff between
several parameters.
 
\keywords {risk perception, SIS model, complex networks}
\end{abstract}

\section{Introduction}
Epidemic spreading is one of the most successful and most studied applications in the field of complex networks. The comprehension of the spreading behavior of many
diseases, like sexually transmitted diseases (i.e. HIV) or the H1N1 virus, can be studied
through computational models in complex networks~\cite{barrat2008,Newman2010}. In addition
to ``real'' viruses, spreading of information or computer malware in technological networks
is of interest as well.
 
The susceptible-infected-susceptible (SIS) model is often used to study the spreading of
an infectious agent on a network. In this model an individual is represented as a node, which can be
either be ``healthy'' or ``infected''. Connections between individuals along which the
infection can spread are represented by links. In each time step a healthy node is infected
with a certain probability if it is connected to at least one infected node, otherwise it
reverts to a healthy node (parallel evolution).

The study of epidemic spreading is a well-known topic in the field of physics and
computer science. The dynamics of infectious diseases has been extensively studied in
scale-free networks~\cite{PhysRevE.63.066117,PhysRevLett.90.028701,PhysRevE.76.061904,PhysRevE.66.036113},
in small-world networks~\cite{PhysRevE.61.5678} and in several kind of regular and random graphs.

A general finding is that it is hard to stop an epidemic in scale-free networks with slow
tails, at least in the absence of correlations in the network among the infections process
and the node characteristics~\cite{PhysRevE.63.066117}. This effect is essential due to the
presence of hubs, which act like strong spreaders. However, by using an appropriate policy
for hubs, it is possible to stop epidemics also in scale-free networks~\cite{stopping,PhysRevE.76.061904}. 

This network-aware policy is inspired by the behavior of real human societies, in which
selection had lead to the development of strategies used to avoid or reduce infections. However,
human societies are not structureless, and a particular focus must be devoted to the
community structures, which are highly important for our social behavior.

Recently, a wave of studies focused the attention on the effect of the community structure
in the modelling of epidemic spreading \cite{10.1371/journal.pcbi.1000736,Chen20121848,in}.
However, the focus was only set towards the interaction between the viruses' features and
the topology, without considering the important relation between cognitive strategies used
by subjects and the structure of their (local) community/neighborhood.

Considering this scenario, an important challenge is the comprehension of the structure of
real-world networks~\cite{communities,Girvan02,fortunato2007}. Given a graph, a community
is a group of vertices that is ``more linked'' within the group than with the rest of the
graph. This is clearly a poor definition, and indeed, in a connected graph, there is not
a clear distinction between a community and a rest of the graph. In general, there is a
continuum of nested communities whose boundaries are somewhat arbitrary: the structure of
communities can be seen as a hierarchical dendrogram~\cite{Newman}.

It is generally accepted that the presence of a community structure plays a crucial role
in the dynamics of complex networks; for this reason, lots of energy has been invested
to develop algorithms for the detection of communities in networks~\cite{Fortunato201075,Modus,MCL}.
However, in complex networks, and in particular in social networks, it is very difficult
to give a clear definition of a community: nodes often belong to more than just one
cluster or module. The problem of overlapping communities was exposed in \cite{Palla}
and recently analyzed in \cite{Lanc}. People usually belong to different communities
at the same time, depending on their families, friends, colleagues, etc. For instance,
if we want to analyze the spreading of sexual diseases in a social environment, it is
important to understand the mechanism that leads people to interact with each other.
We can surely detect two distinct groups of people (\textit{i.e.}, communities):
heterosexual and homosexual, with bisexual people that act as overlapping vertexes
between the two principal communities~\cite{citeulike:3856997,gcalda:AM92,Chen20121848}. 

The strategies used to face the infection spreading in a community is itself a
complex process (\textit{i.e.}, social problem solving) in which strategies spread
(as the epidemics) along the community, and are negotiated and assumed or discarded
depending on their social success.

Several factors can affect the social problem solving which is represented by 
the adoption of a behaviour to reduce the infection risk. Of course,
personality factors, previous experiences and the social and economical states
of a subject can be considered as influencing variables. Another important
variable is represented by the structure of the environment in which the social
communities live, because it determines at the same time the speed of the
epidemic diffusion and the strategy of the negotiation process; in particular large
and more connected communities are often characterized by conservative
strategies while small and isolated communities allow more relaxed strategies.

The same strategy can be more or less effective depending on the strategies
adopted by the neighbours (community) of the subject. For instance, a subject
in a conservative community can adopt a more risky (and presumably profitable)
attitude with a certain confidence since he would be protected from the
infection, because of their neighbours' behaviours, and this ``parasitic''
behavior (like refusing vaccinations) can be tolerated (up to a certain level)
without lowering the community's fitness.

Not only the neighbor's behaviour affect the evolution
of the cognitive strategies of a subject, but also the position he has in the network
should be a relevant factor. A hub, or a subject with a great social betweenness,
is usually more exposed to the infection than a leaf, and as a consequence,
the best strategy for him has to be different. In the same way, since the
topology of the network (\textit{e.g.}, small world, random) determines
variables such as the speed of the spreading, or its pervasiveness, it should
also affect the development of the ``best strategy''. 

Moreover, while the negotiation process evolves, the cognitive strategies usually
develop within the most intimate community of a subject, thus the behaviour
adopted by subjects could be an interesting feature for the community detection
problem as well.

The understanding of the effects of the community structure on the epidemic
spreading in networks is still an open task. In this paper we investigate the role
of risk perception in artificial networks generated in order to reproduce
several types of overlapping community structures.

The rest of this paper is organized as follows: we start by describing a mechanism
for generating networks with overlapping community structures in section~\ref{level2}.
In section~\ref{terzo}, we describe the SIS model adopted to model the risk
perception of subjects in those networks. Finally, section~\ref{quarto} contains
simulation results from our model with a troughout discussion and future work proposals.

\section{The networks model}
\label{level2}

\begin{figure}[t!]
\centering
\subfigure[] 
{\includegraphics[width=0.47\columnwidth]{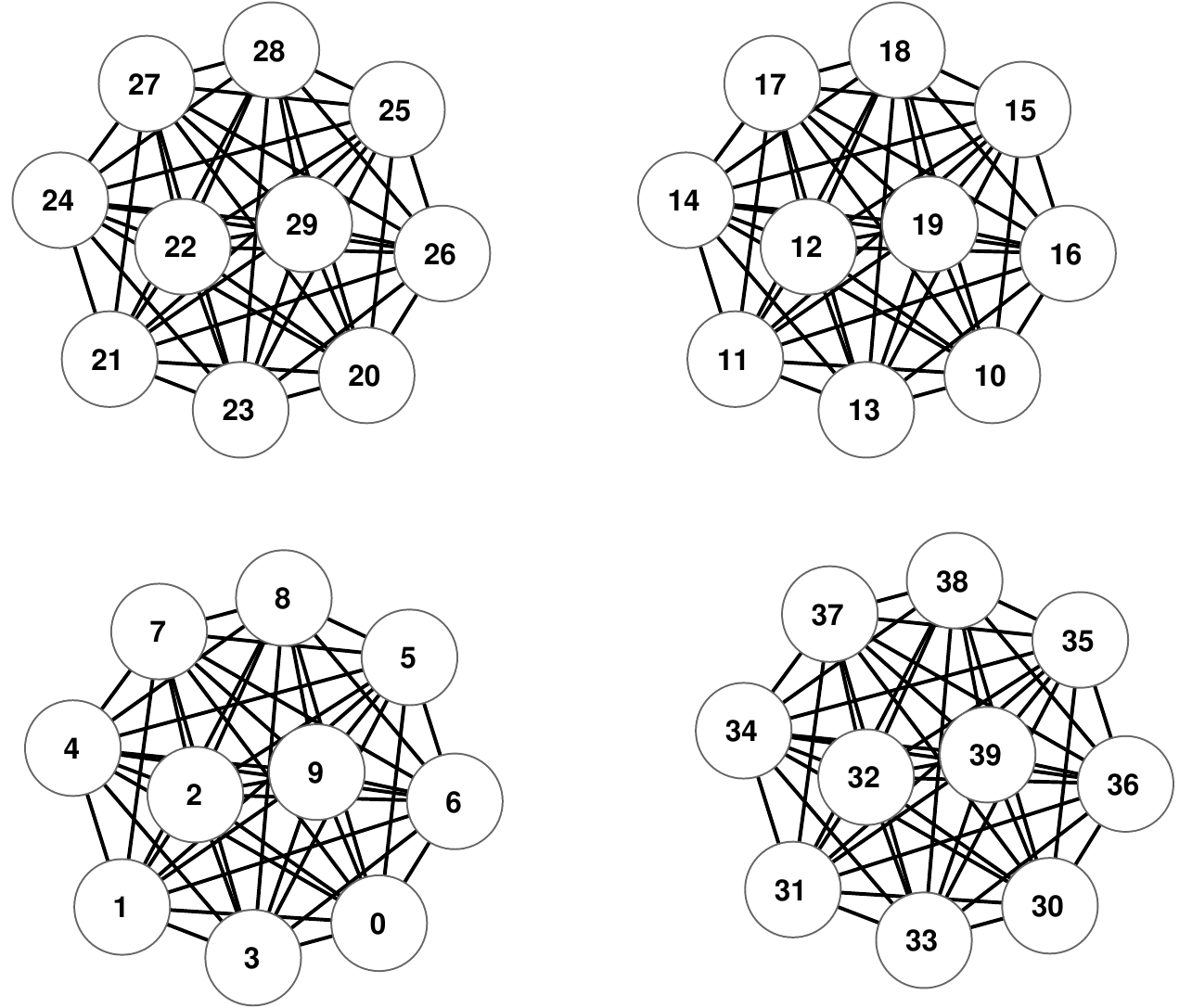}}
\hspace{2mm}
\subfigure[]
{\includegraphics[width=0.47\columnwidth]{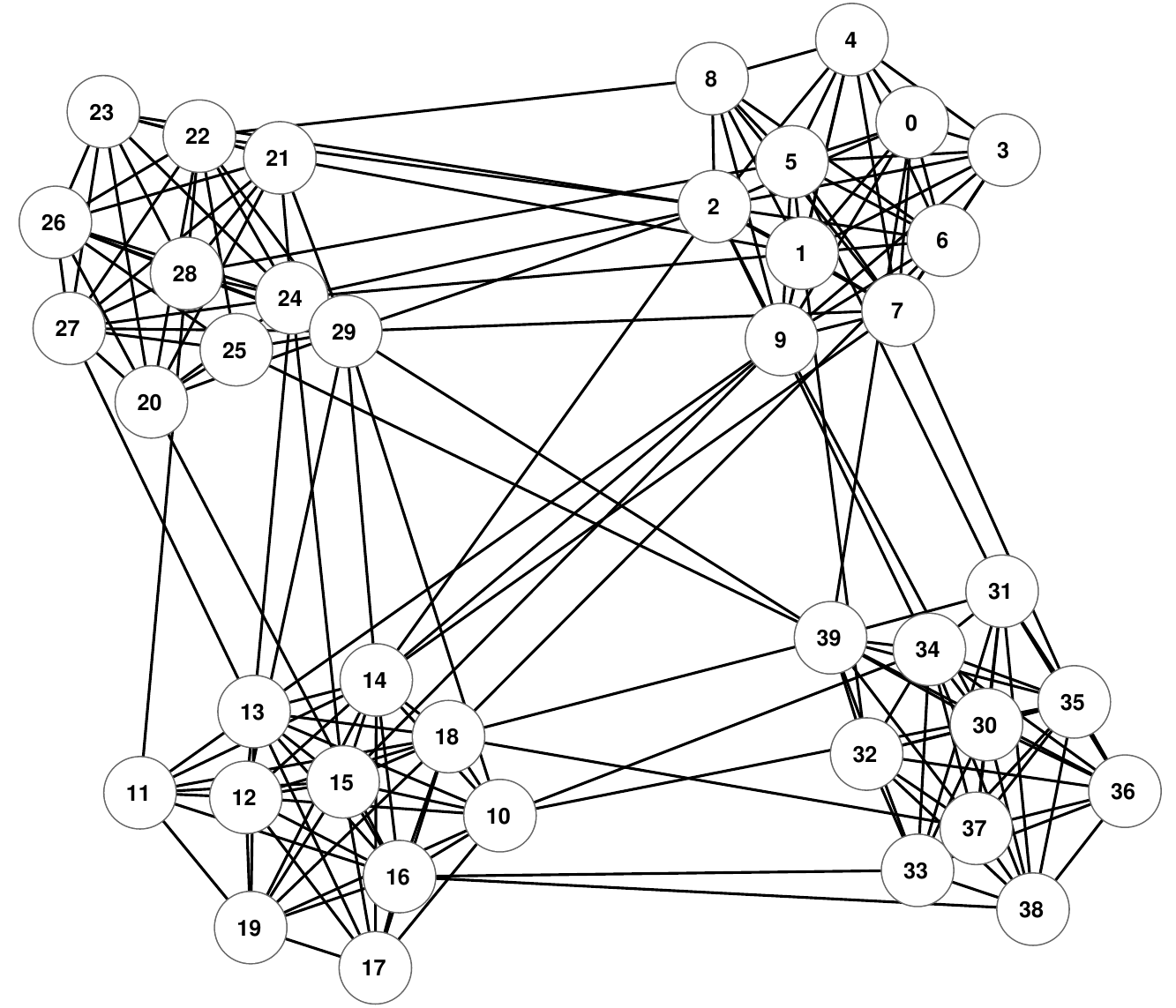}}
\caption{\label{fig:net} (a) An example network with $4$ different
communities composed by $10$ vertices: in this case, considering
$p_1=1$ and $p_2=0$, we generate $4$ non interconnected full
connected networks. (b) The same $4$ communities with parameters
$p_1=0.95$ and $p_2=0.05$.}
\end{figure}

There are $n_c$ different communities with $n_v$ vertices (in this
paper we consider only undirected and unweighted graphs); we assume
that the probability to have a link between the vertexes in the same
community is $p_1$, while $p_2$ is the probability to have a link
between two nodes belonging to different communities. For instance,
with $p_1=1$ and $p_2=0$, we generate $n_c$ fully connected graphs,
with no connections among them as shown in figure \ref{fig:net}(a).
It is possible to use the parameters $p_1$ and
$p_2$ to control the interaction among different communities, as
shown in \figurename~\ref{fig:net}(b). The algorithm for generating
this kind of networks can be summarized as:

\begin{enumerate}
\item Define $s_1$ as number of vertexes in the communities;
\item Define $n_c$ as number of communities;
\item For all the $n_c$ communities create a link between the
	vertexes on them with probability $p_1$;
\item For all the vertexes $N = s_1 n_c$ create a link between
	them and a random vertex of other communities with probability $p_2$;
\end{enumerate}

\begin{figure}[t!]
\centering
\subfigure[] 
{\includegraphics[width=0.495\columnwidth]{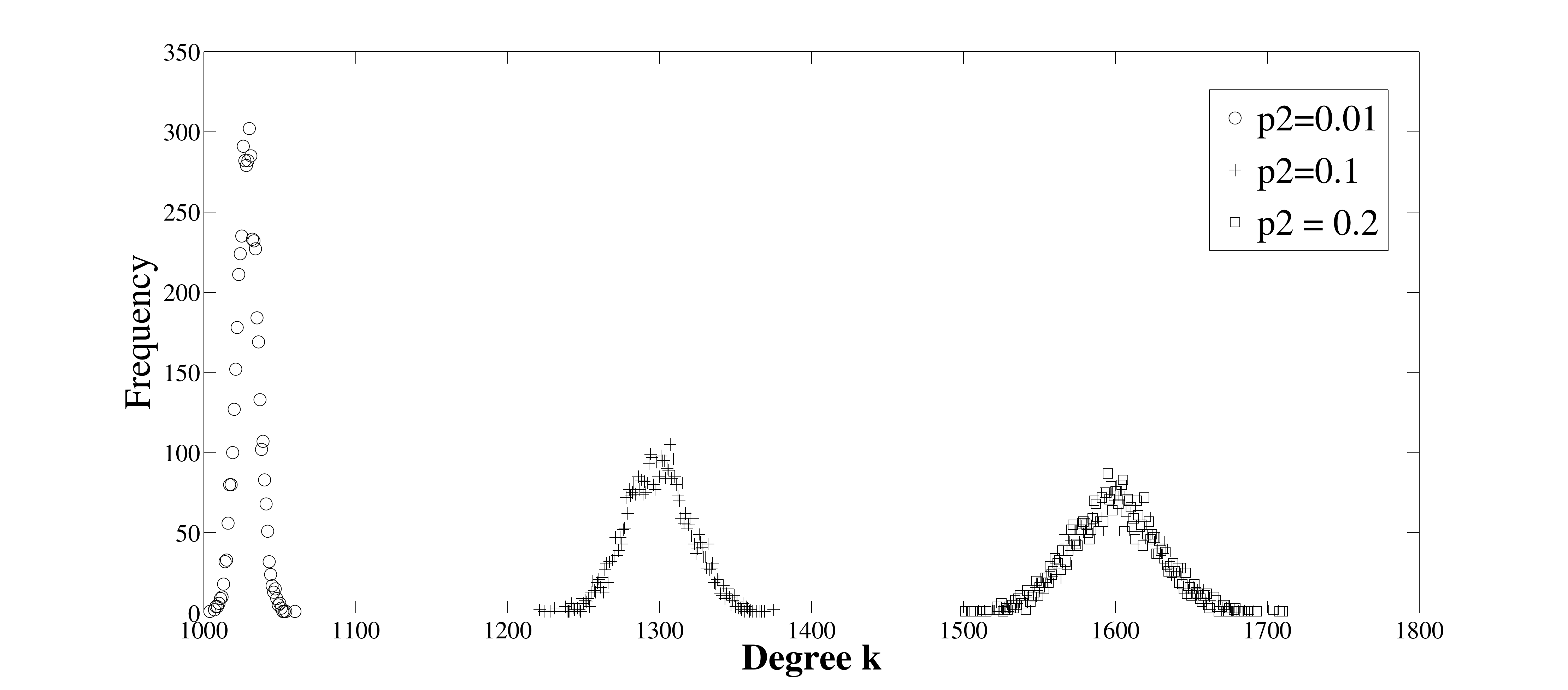}}
\subfigure[]
{\includegraphics[width=0.495\columnwidth]{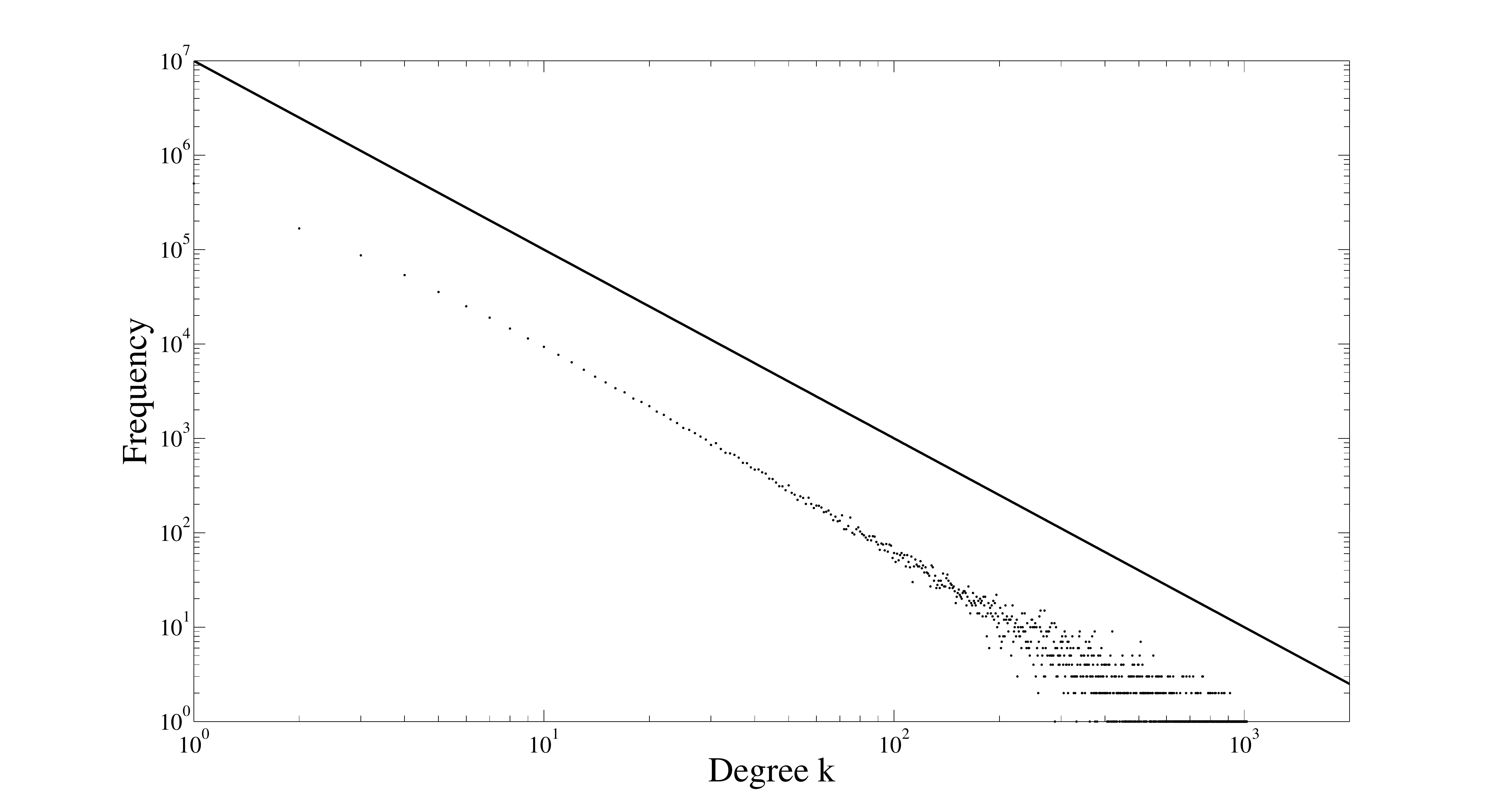}}
\caption{\label{fig:degree} (a) Random networks: in this figure, we show the frequency
distribution of the connectivity degree changing the value of the
parameter $p_2$. The circles represent the values for $p_2=0.01$,
crosses for $p_2 = 0.1$ and eventually squares for $p_2=0.2$.
Here, $s_1 = 1000$ and $n_c = 5$, thus we have generated networks
with $5$ communities of $1000$ nodes for each. (b) Distribution of
connectivity degree for the scale-free network generated with the
mechanism described above (dots). The straight line is a power
law curve with exponent $\gamma = 2.5$.}
\end{figure}

Constraining the condition $p_1=1-p_2$, we can reduce the free parameters to just one.
The connectivity degree itself depends on the size of the network and on the
probabilities $p_1$ and $p_2$. 
In particular the connectivity function $f(k)$ has a normal distribution from which we could define the mean connectivity $\left \langle k \right \rangle$, as:
\begin{equation}
\left \langle k \right \rangle = (s_1 - 1)p_1 + (n_c - 1)s_1p_2
\label{mean}
\end{equation}
with standard deviation $\sigma^2(k) = (s_1-1)p_1(1-p_1) + (n_c-1)s_1p_2(1-p_2)$.

In figure~\ref{fig:degree}(a) we show the frequency
distribution of the connectivity degree of nodes varying the value of the parameter
$p_2$ for a network composed by $N = 5000$ nodes and $n_c=5$ communities.

It is widely accepted that real-world networks, from social networks to computer networks
are scale-free networks, whose degree distribution follows a power law, at least
asymptotically. In this network, the probability distribution of contacts often exhibits
a power-law behavior:

\begin{equation}
P(k) \propto ck^{-\gamma},
\end{equation}

with an exponent $\gamma$ between $2$ and $3$~\cite{sf,PhysRevLett.85.4633}. For generating
networks with this kind of networks we adopt the following mechanism:

\begin{enumerate}
\item Start with a fully connected network of $m$ nodes;
\item Add $N-m$ nodes;
\item For each new nodes add $m$ links;
\item For each of these links choose a node at random from the ones already belonging to
      the network and attach the link to one of the neighbors of that
      node, if not already attached.
\end{enumerate}

Through this mechanism we are able to generated scale-free networks with an
exponent $\gamma{}=2.5$ as shown in figure~\ref{fig:degree}(b). There, we show the
frequency distribution of the connectivity degree for a network of $10^6$ nodes. 
To generate a community structure with a realistic distribution, we first
generate $n_c$ scale-free networks as explained above. Then, for all nodes and all
outgoing links, we replace the link pointing inside the community with that connecting
a neighbor of a random node in a random community with a probability of $p_2=1-p_1$.
Thus, the algorithm can be summarized as:

\begin{enumerate}
\item Generate $n_c$ communities as scale-free networks with $n_v$ vertices;
\item For all the vertices, with a probability $p_2=1-p_1$;
\begin{itemize}
\item Delete a random link;
\item Select a random node of another community and create a link with one of its adjacent vertex;
\end{itemize}
\item End.
\end{enumerate}

\begin{figure}[t!]
\centering
{\includegraphics[width=0.7\columnwidth]{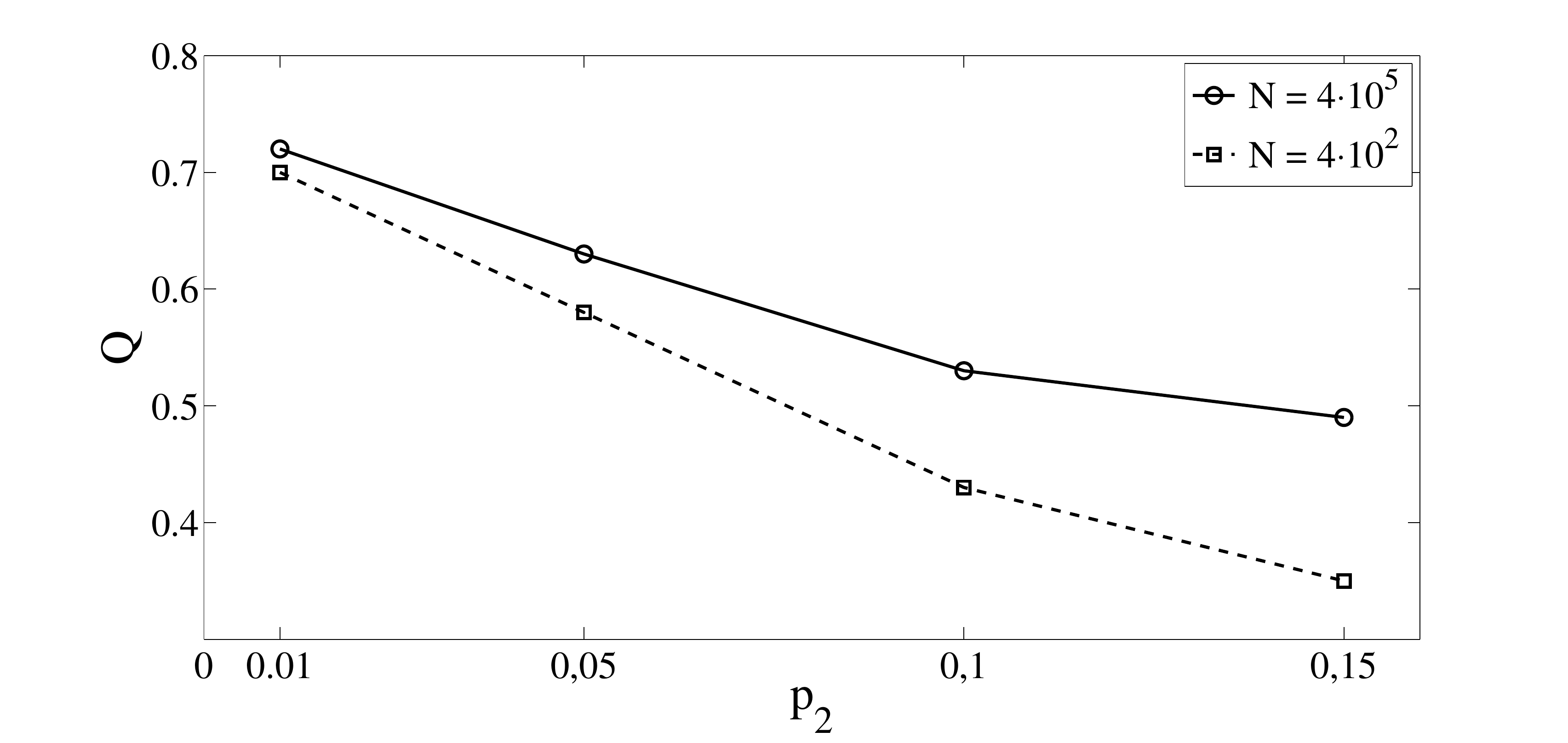}}
\caption{\label{fig:mod} Different values of modularity ($Q$) after increasing the mixing
parameter $p_2$ for two networks with $N = 4\cdot 10^5$ nodes and $N = 4\cdot 10^2$ nodes.}
\end{figure}

In this way, we are able to generate scale-free networks with a well defined community
structure. A good measure for the estimation of the strength of the community structure
is the so-called modularity~\cite{Girvan02}.
The modularity $Q$ is defined to be:

\begin{equation}
Q = \frac{1}{2}\sum_{vw}\left[A_{vw}-\frac{K_v K_w}{2m}\right]\delta(c_v,c_w),
\end{equation}

where $A$ is the adjacency matrix in which $A_{vw}=1$ if $w$ and $v$ are connected
and $0$ otherwise. 
$m=\frac{1}{2}\sum_{vw}A_{vw}$ is the number of edges in the graph, $K_i$ is the
connectivity degree of node $i$ and $(K_vK_w)/(2m)$ represents the probability of an
edge existing between vertices $v$ and $w$ if connections are made at random but with
respecting vertex degrees. $\delta(c_v,c_w)$ is defined as follows:

\begin{equation}
\delta(c_v,c_w)=\sum_r^{n_c} \hat{c}_{vr} \hat{c}_{wr}
\end{equation}

where $\hat{c}_{ir}$ is $1$ if vertex $i$ belongs to group $r$, and $0$ otherwise.

In figure~\ref{fig:mod} we show the values of modularity for two networks that were
generated with the same algorithm, but with different sizes. Here, we consider a
network with $4$ communities: in the first case $s_1=10^5$, while in the second case
$s_1=10^2$. What one can observe in figure~\ref{fig:mod} is that the modularity's
behaviour does not change significantly for different network sizes with the same number of communities.

In the case of scale-free networks, the mean connectivity degree $\left \langle k \right \rangle$ is fixed
a priori when we choose the number of links the new nodes create. In the case
of random networks the mean connectivity is given by equation~\ref{mean}.

\section{The risk perception model}
\label{terzo}
We use the susceptible-infected-susceptible model (SIS)~\cite{PhysRevE.63.066117, gcalda:AM92}
for describing an infectious process. In the SIS model, nodes can be in two distinct states:
healthy and ill. Let us denote by $\tau$ the probability that the infection can spread along
a single link. Thus, if node $i$ is susceptible and it has $k_i$ neighbors, of which $s_{n}$
are infected, then, at each time step, node $i$ will become infected with the probability:
\begin{equation}
p(s,k)=[1-(1-\tau)^{s_{n}}].
\label{probability}
\end{equation}
We model the effect of risk perception considering the global information of the infection
level for the whole network, the information about the infected neighbors and the information
about about the average state of the community. Thus, the risk perception for the individual
$i$ is given by:

\begin{equation}
I_i=\exp\left\{-H+J_1\left(\frac{s_{ni}}{k_i}\right)+J_2\left(\frac{s_{ci}}{n_{ci}}\right)\right\},
\label{perc}
\end{equation}

where $H=J(s/N)$ is the perception about the global network on which $s$ is the total number
of infected agents while $N$ is the number of agents in the network. The second term of the
equation~\ref{perc} represents the perception about the neighborhood, while the third term
represents the perception about the local community of the agent $i$.

In this model, we assume that people receive information about the network's state through
examination of people in the neighborhood. The global information could refer to entities like
media while the information about the community could be assumed as \emph{word of mouth}.
In this paper, we don't consider the cost that people should pay in order to get these
information, but it is clearly an important constraint to consider in future works.

The risk perception $I_i$, defined in equation~(\ref{perc}), is assumed to determine the
probability that the agents meet someone in its neighbourhood. The algorithm is given by:

\begin{enumerate}
\item For all nodes $i=1,2,\dots,N$;
\item For all its neighbors $j=1,2,\dots,k_i$;
\item If $I_i>rand$; 
\begin{itemize}
\item $i$ meets $j$;
\item If $j$ at time $t-1$ was infected then $i$ becomes ill with probability $\tau$; 
\end{itemize} 
\item End.
\end{enumerate}
Then, we propose a gain function defined as the number of meetings in time considering
different values of $j=J,J_1,J_2$ and different kind of scale-free and random networks;
the gain function $G(j)$ is given by:

\begin{equation}
G(j)=\frac{\sum_{t=1}^{T_e}M_t}{T_e},
\end{equation}

in which $T_e$ is the time for the extinction, while $M_t$ is the number of meetings
during time. Based on that, we can eventually define a fitness function that considers
the probability to extinct the epidemic in the given time. Thus, the fitness function is
given by:

\begin{equation}
F^T_j=G(j)P_e(j)
\end{equation}

It is possible to make a mean-field approximation of this model. Pastor-Satorras and
Vespignani defined the mean-field equation for scale-free networks in \cite{PhysRevE.63.066117}.
In 2010, Kitchovitch and Li\`{o}~\cite{citeulike:7329585} modeled the mean number
of infected neighbors $g(k)$ for individuals $i$ with connectivity degree $k$. In fact,
given the probability of receiving an infection by at least one of the infected neighbors
(equation~\ref{probability}), it is possible to define the rate of change of the
fraction of individuals $i$ with degree $k$ at time $t$ by the following:
\begin{equation}
\frac{di_k}{t} = -\gamma + (1-i_k)g(k),
\label{rate}
\end{equation}
on which $\gamma$ is the rate of recovery (in our simulations we set $\gamma=1$). 

Then, as shown by Boccaletti et al~.\cite{boccaletti06}, for any node, the degree
distribution of any of its neighbors is,
\begin{equation}
q_k=\frac{kP(k)}{\left \langle k \right \rangle},
\end{equation}
hence it is possible to define the number of infected neighbors as:
\begin{equation}
i_n = \sum_{K_{min}}^{k_{max}}q_ki_k,
\end{equation} 
and it allows to give a definition of $g(k)$, as:
\begin{equation}
g(k) = \sum_{s=0}^{k}{k \choose s} p(s,k)i_n^s(1-i_n)^{k-s},
\end{equation}
where $s=s_n$ is the number of infected neighbors.

The temporal behavior of the mean fraction $c$ of infected individuals
in the case of a network with fixed connectivity is given by:

\begin{equation}
c^{'}=\sum_{s_{n}=1}^k{k\choose{}s_{n}}c^{s_{n}}(1-c)^{k-s_{n}}[1-(1-\tau)^{s_{n}}],
\end{equation}

where $c\equiv{}c(t)$, $c'\equiv{}c(t+1)$ and the sum runs over
the number $k_{inf}$ of infected individuals.

\begin{figure}[h!]
 \centering
 \subfigure[]
   {\includegraphics[width=.45\columnwidth]{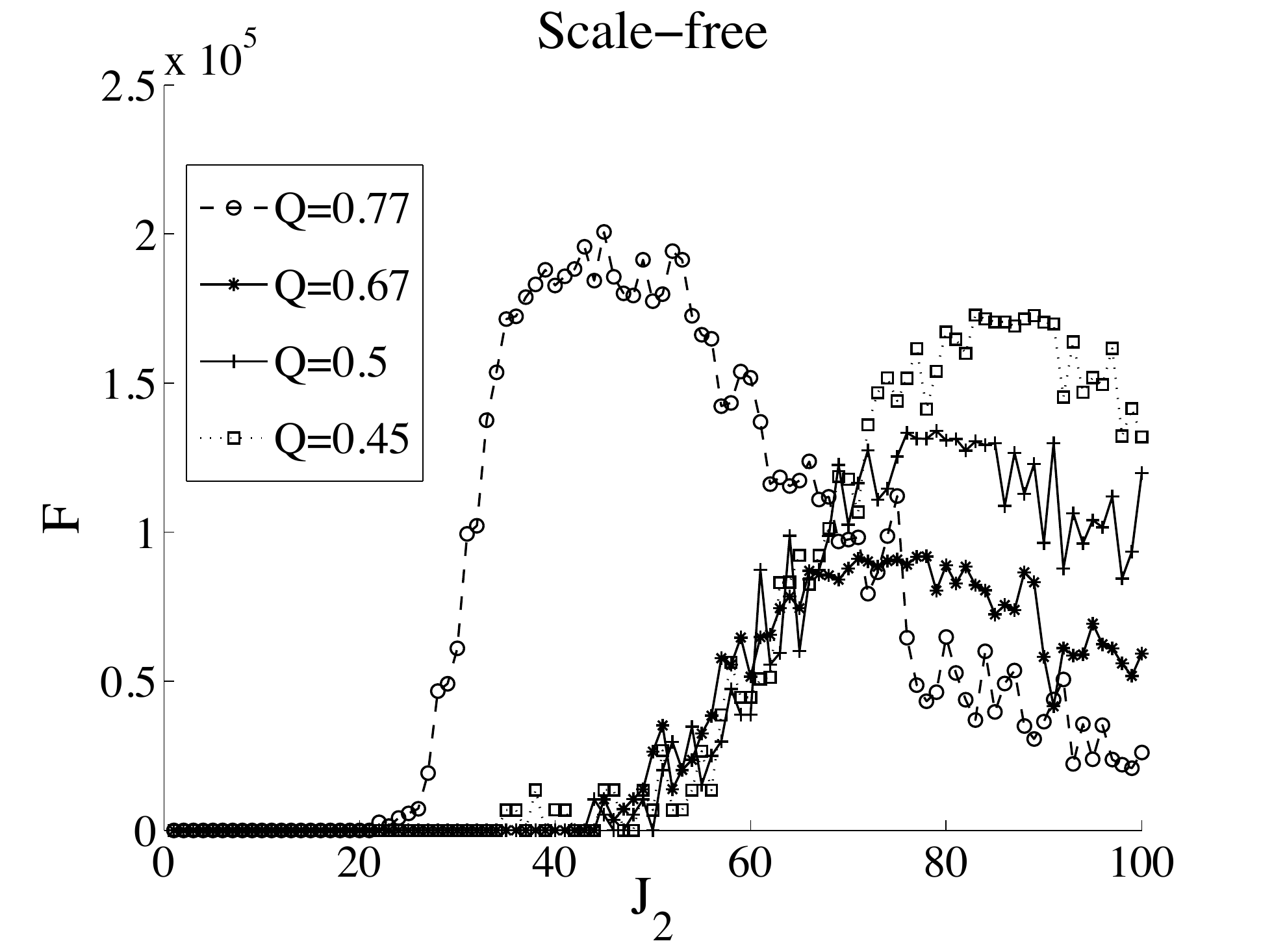}}
 \hspace{1mm}
 \subfigure[]
   {\includegraphics[width=.45\columnwidth]{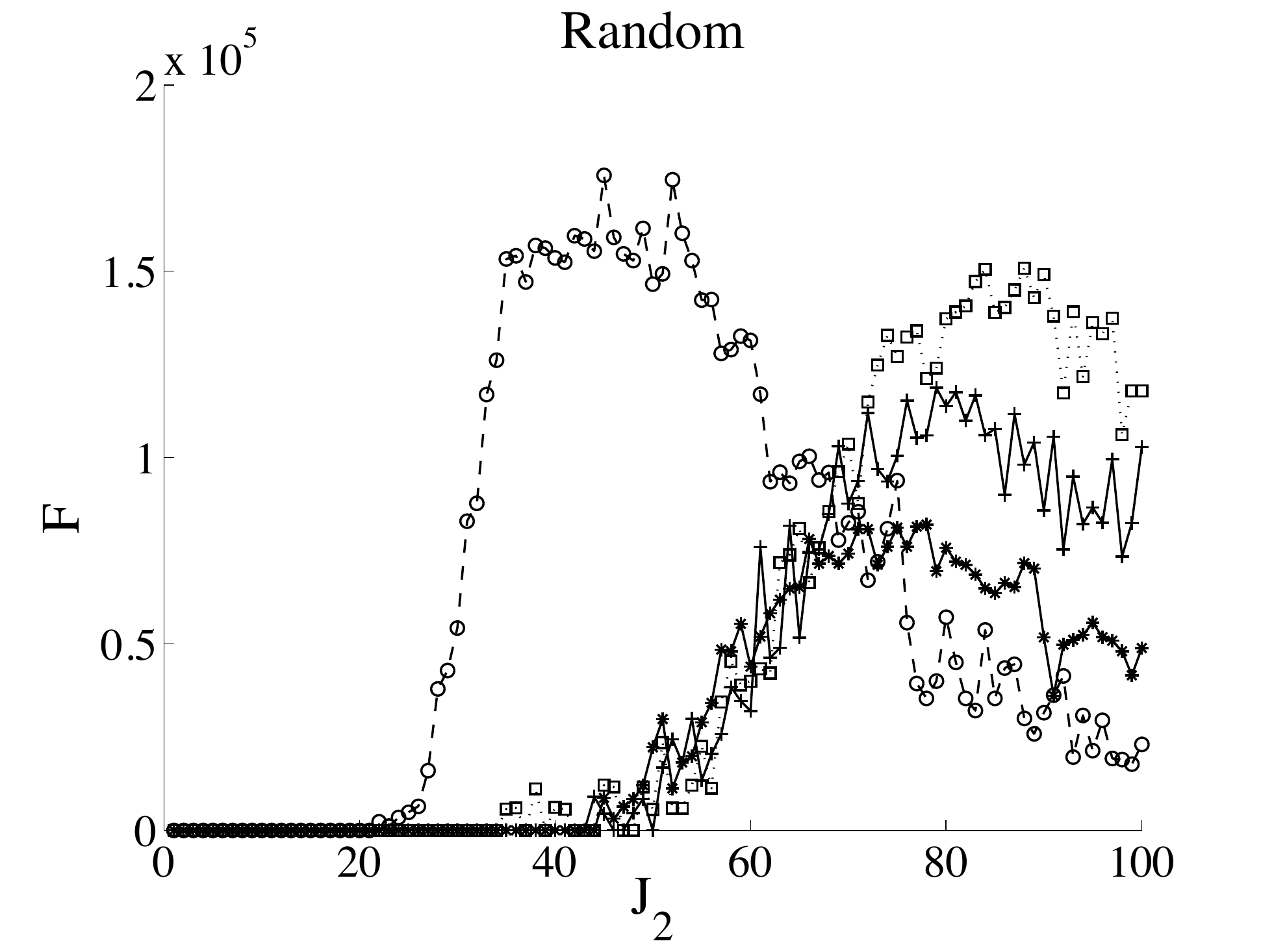}}
   \hspace{1mm}
 \subfigure[]
   {\includegraphics[width=.45\columnwidth]{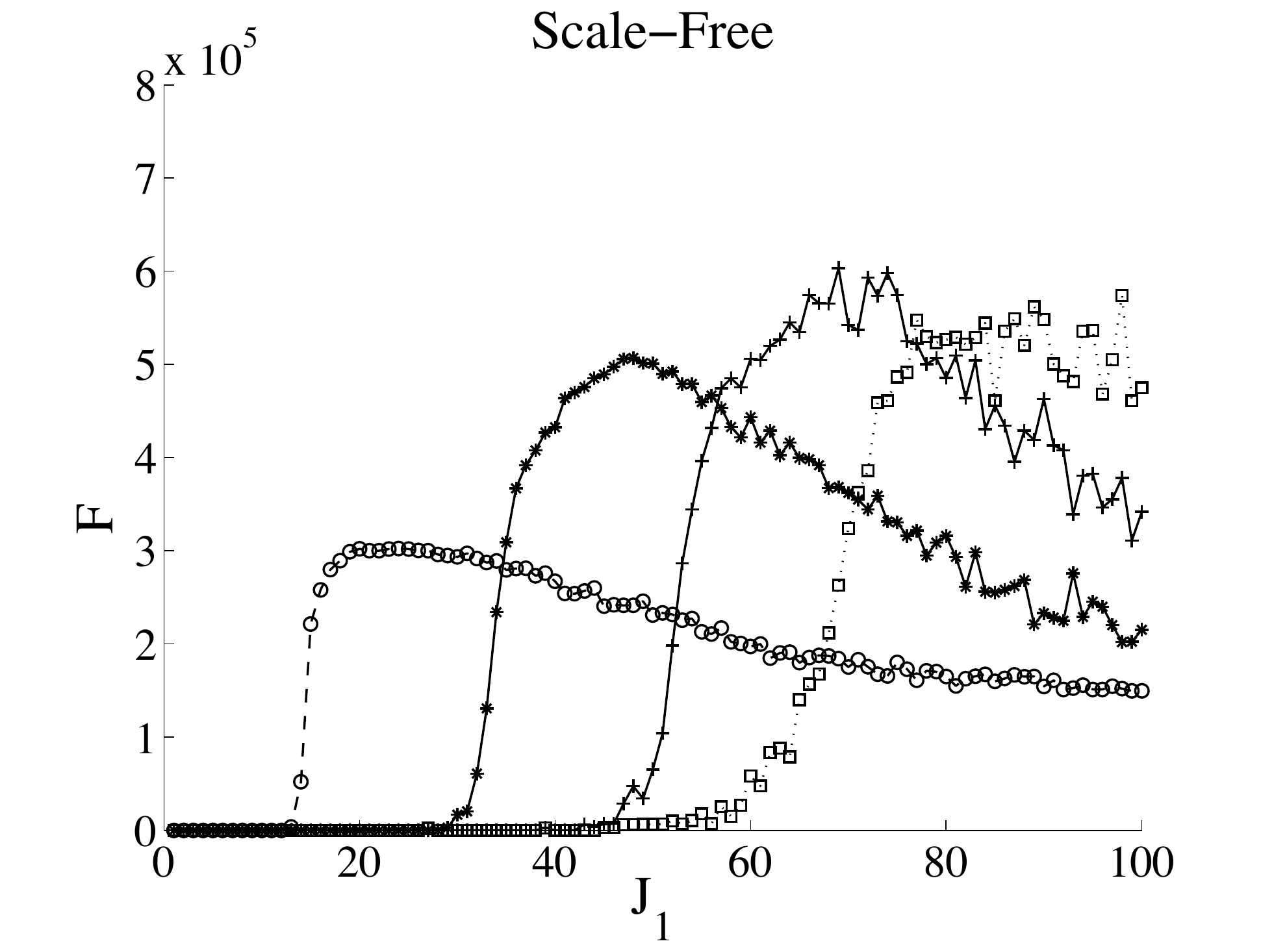}}
      \hspace{1mm}
 \subfigure[]
   {\includegraphics[width=.45\columnwidth]{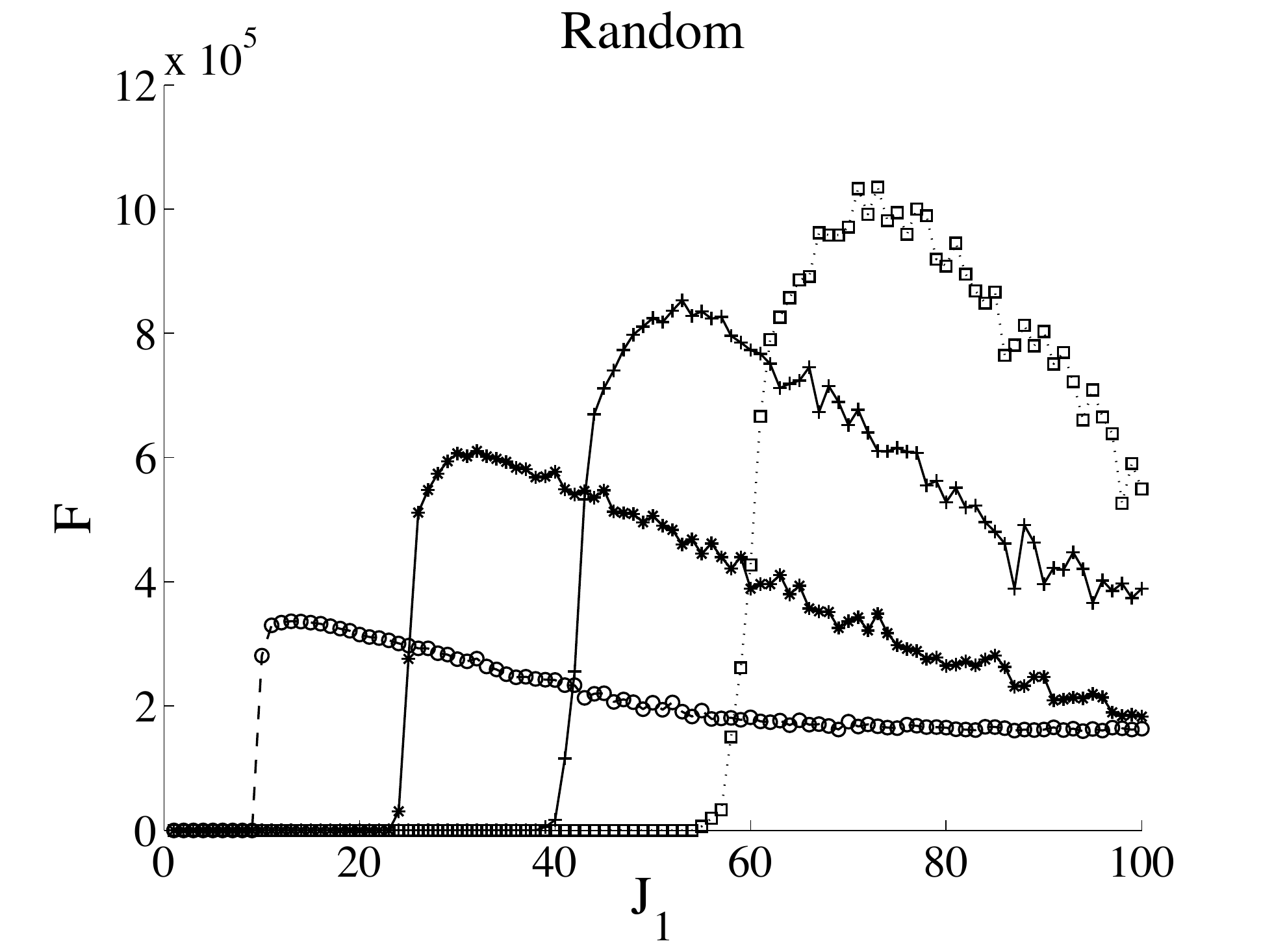}}
      \hspace{1mm}
 \subfigure[]
   {\includegraphics[width=.45\columnwidth]{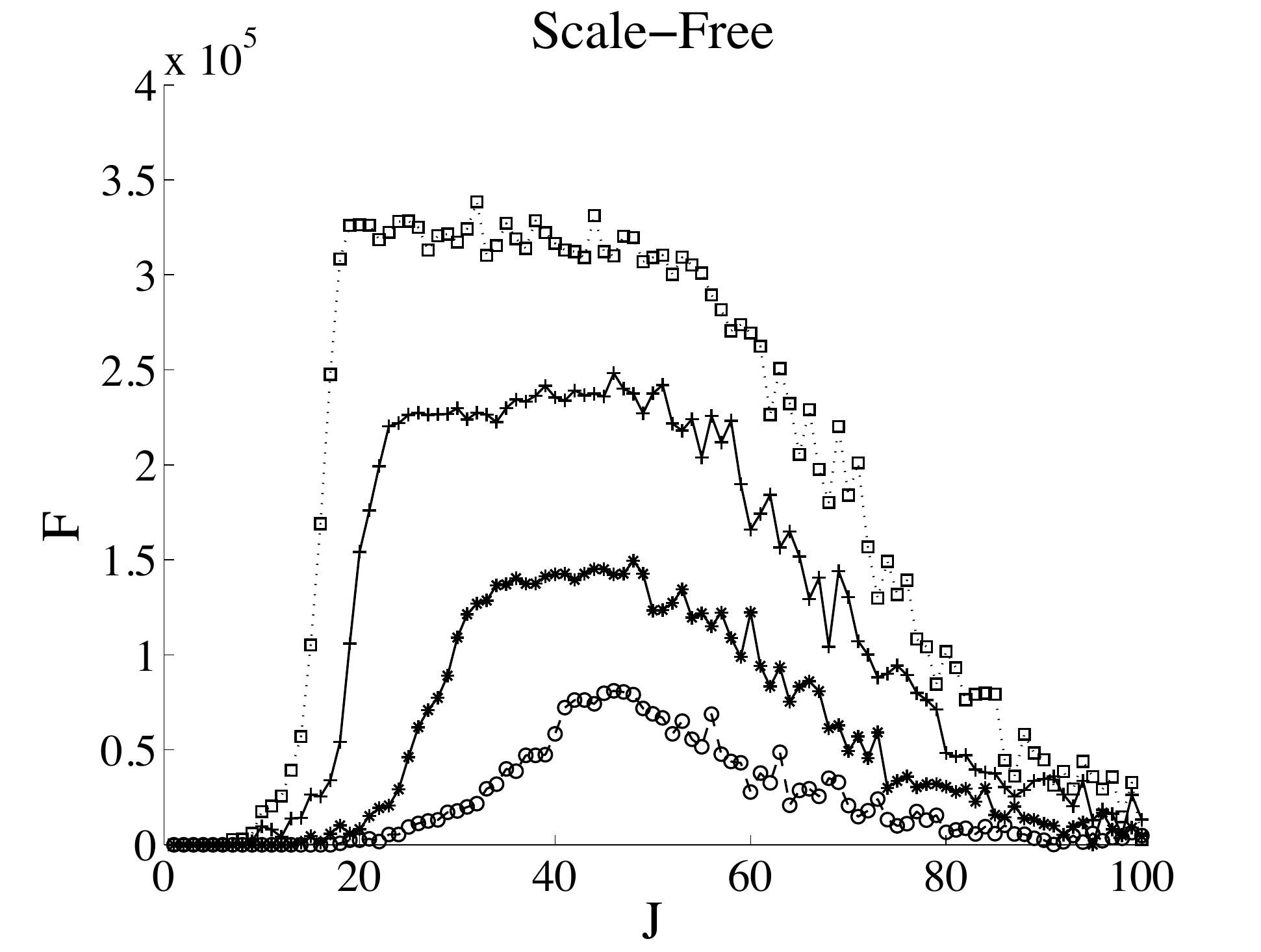}}
      \hspace{1mm}
 \subfigure[]
   {\includegraphics[width=.45\columnwidth]{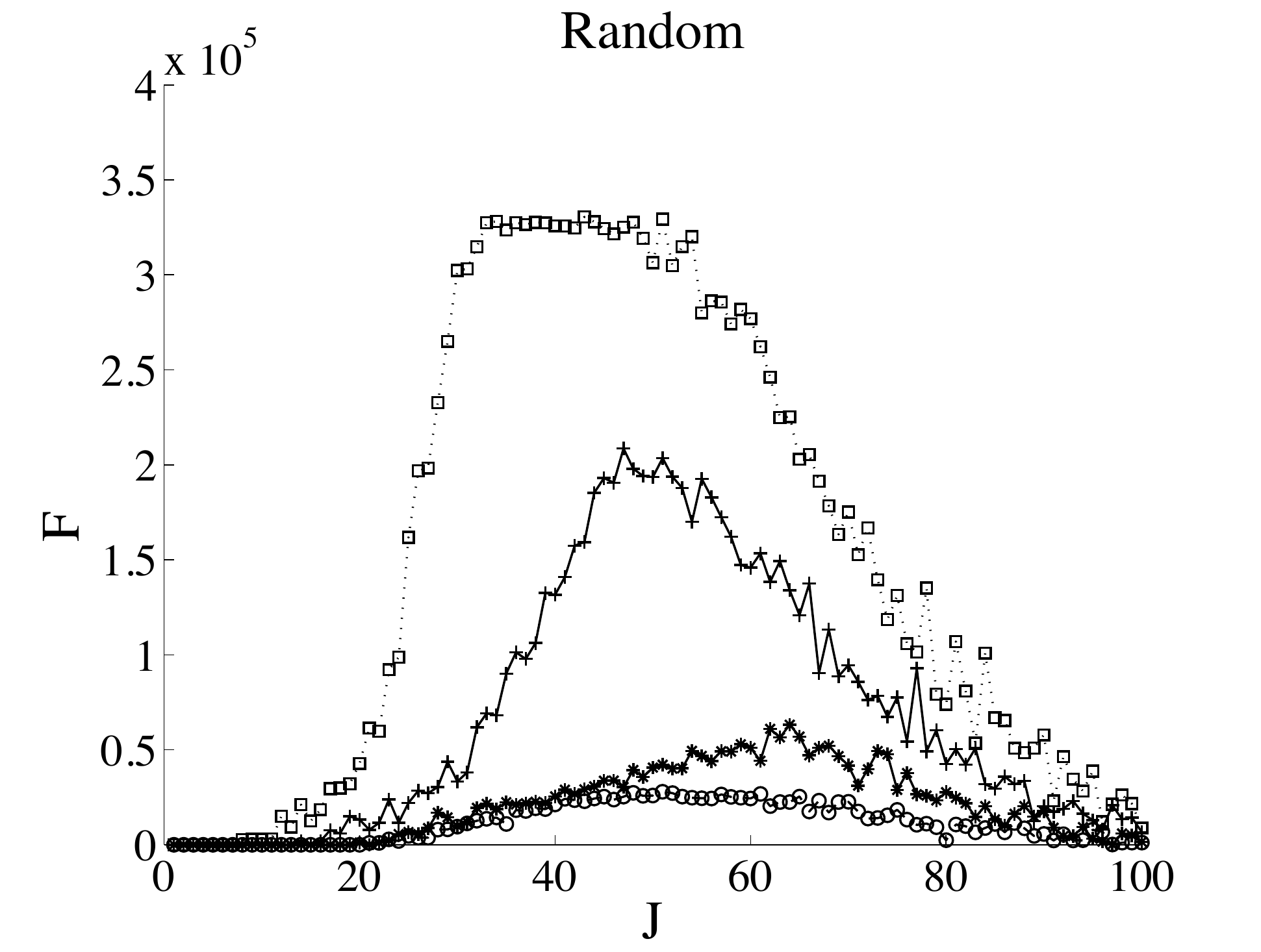}}
\caption{\label{fig:res} Effect of the parameters  $J_2$~(a), $J_1$~(b) and $J$~(c)
(\emph{x-axis}) on the fitness function $F$ considering different scale-free
and random networks with different values of modularity. Results are averaged over
$100$ simulations for each value of $J$ (but actually $J$, $J_1$ and $J_2$).}
\end{figure}

\section{Results and Discussion}
\label{quarto}
\begin{figure}[t!]
\centering
{\includegraphics[width=1\columnwidth]{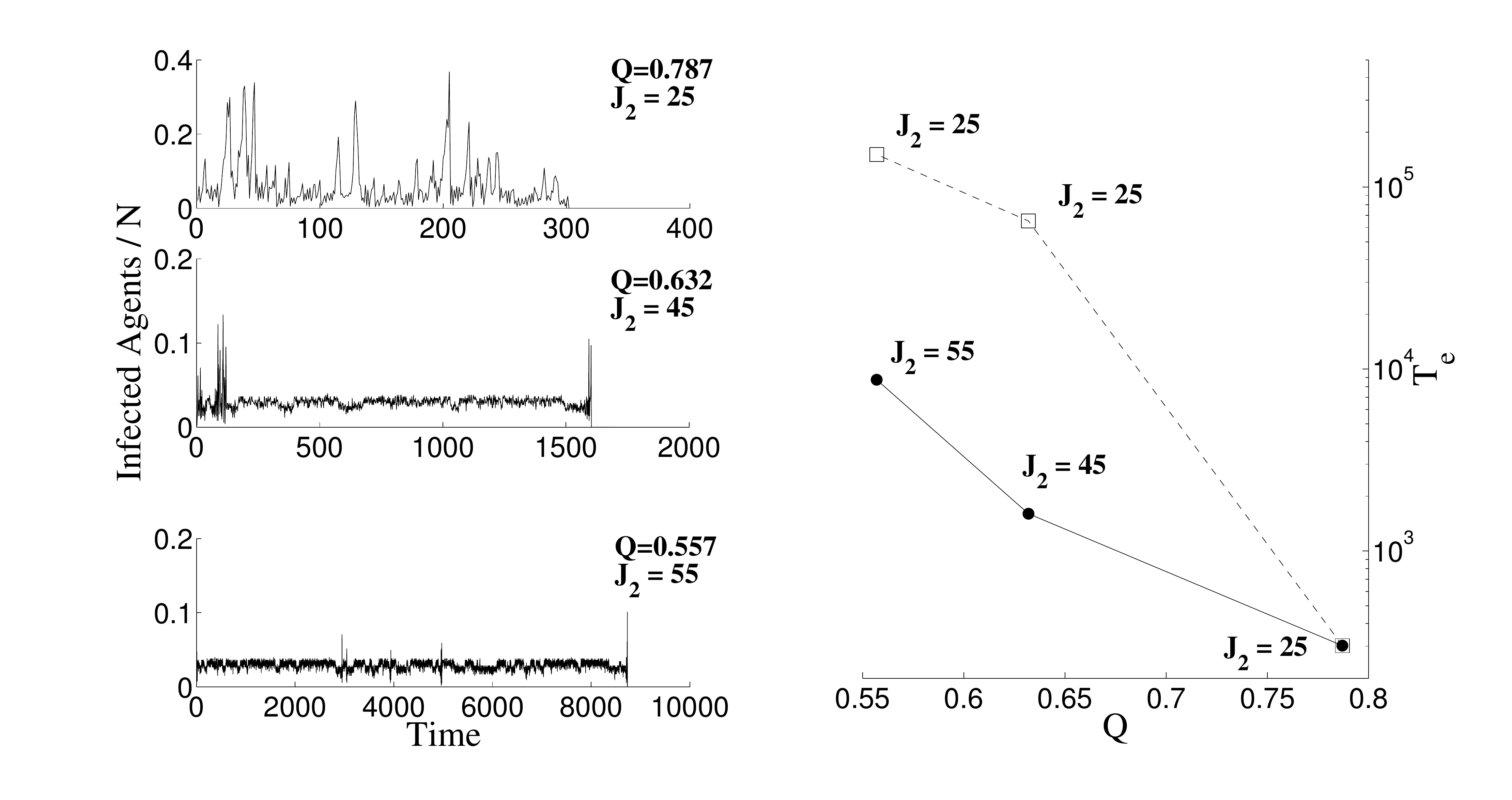}}

\caption{\label{fig:res1} On the left side of the figure we show the temporal evolution of
infected individuals by varying the mixing parameter $p_2$. The time necessary for the
epidemic extinction increases as the modularity $Q$ decreases. On the right side, we show
the effects of the precaution parameter $J_2$ on the extinction time by varying the
modularity $Q$. The straight line represents the results for different value of $J_2$,
while the dashed line represents the results for a constant value of $J_2$.}
\end{figure}

We studied the behavior of our model for different scenarios. In figure~\ref{fig:res1},
we show results considering a network of $500$ nodes and $5$ communities where the initial
number of infected agents is $\approx{}10\%$ of all agents in the network. We
focus on the information about the community (parameter $J_2$), while we kept fixed
$J=J_1=1$. It is very interesting to observe the time necessary for the extinction of the
epidemics, with the probability of being infected $\tau=0.5$ and changing the community
structure of the network. 

On the left side of figure~\ref{fig:res1} we show the temporal evolution of the percentage
of infectious agents for different kind of networks and different values of $J_2$.
We can observe that the extinction time increases when the modularity of network decreases,
even if we use higher values of $J_2$. On the right side of figure~\ref{fig:res1}, we show
the effect of the precaution on the extinction time. The straight line corresponds to
different values of $J_2$, while the dotted line corresponds to the same value of $J_2$
in different kind of networks. It is also possible to observe that when a network becomes
\emph{less clustered}, the information about the community becomes less important.

In the case of scale-free networks, the mean connectivity degree $\left \langle k \right \rangle$ is related to
the number $m$ of links the new nodes create. In the above example, considering $m=5$,
we obtained a mean connectivity degree $\left \langle k \right \rangle=7.8$.

\begin{figure}[htb!]
\centering
{\includegraphics[width=0.6\columnwidth]{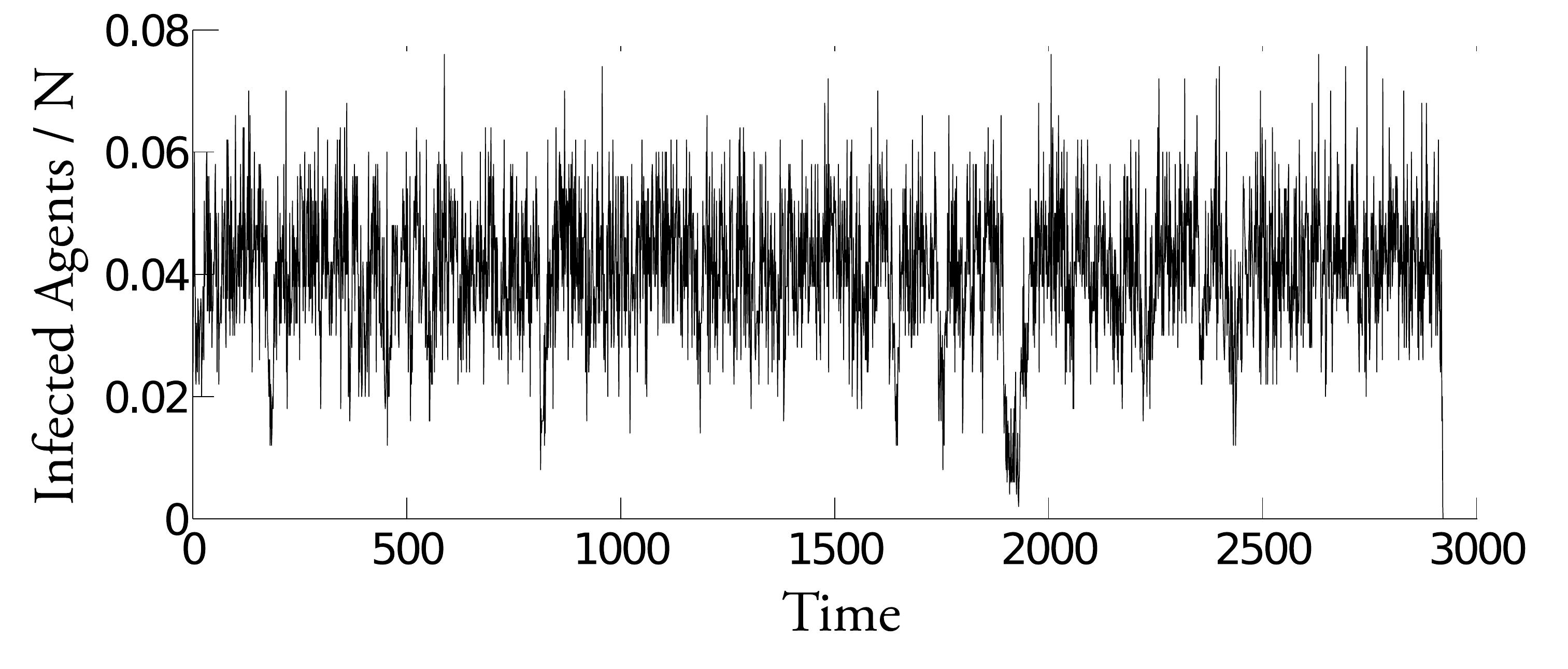}}
\caption{\label{fig:res2} Percentage of infected agents for a random network of $100$ nodes and
$5$ communities with modularity $Q=0.78$. Adopting the same parameter used for the simulation
reported in figure~\ref{fig:res1}, we show how the time for the extinction (approx. $2900$ units) of the epidemic is
greater than for the scale-free case (\textit{i.e.}, upper plot on the left side of
figure~\ref{fig:res1}).}
\end{figure}

For comparisons, we generated random networks with a mean connectivity degree $\left \langle k \right \rangle\in(7,8)$.
The first result that we obtained is that the extinction time is larger than in the scale-free
case. In figure~\ref{fig:res2}, we show the temporal evolution of the infected agents for a
random network with modularity $Q=0.78$ considering $J_2=25$ as in the upper plot on the left
side of figure~\ref{fig:res1}. For the scale-free network the time necessary for the extinction
is $T_e\simeq{}3\cdot10^2$ while for the random one it is $T_e\simeq{}3\cdot10^3$.

\begin{table}
\centering
\begin{tabular}{c|c c c c c}
\hline
\hline
 &  & & Critical Values & & \\
 $Q$ (modularity) & $J$ & & $J_1$ & & $J_2$ \\
\hline
\hline
0.78 & 45 & & 15 & & 25  \\
0.64 & 40 & & 15 & & 45 \\
0.35 & 40 & & 20 & & 55 \\
\hline
\end{tabular}
\caption{Critical values for the extinction of the epidemic on case of scale-free networks of $500$ nodes and $5$ communities considering a  maximum threshold
time $T_{max} = 1000$, necessary for the extinction of the epidemics.}
\label{tab:tab}
\end{table}

Regarding the effects of the global and local (neighborhood) information, we investigated
scale-free networks composed by $500$ nodes and $5$ communities with a fixed maximum threshold
time $T_{max}$, necessary for the extinction of the epidemics. We assume $T_{max}=1000$ and
separately measure critical values of $J,J_1$ and $J_2$. In the table ~\ref{tab:tab}, we show the
critical values of the three parameters by changing the modularity $Q$. As we can observe, the
most variable parameter is $J_2$ while the other two parameters do not appear to change. From
this figure, we observe that the information about the fraction of infected neighbors is the
most effective for stopping the disease. However, in order to get this piece of information, each node
needs to check the status of all its neighbours, a task that can be quite hard and possibly
conflicting with privacy. On the other hand, the information on the average infectious level in
the community or in the whole population is more easily obtained. Therefore, one needs to add
the cost of information into the model in order to decide what the most effective solution
for risk perception is.


Concluding, we have studied the progression and extinction of a disease in a SIS model over modular
networks, formed by a certain number of random and scale-free communities. The infection
probability is modulated by a risk perception term (modeling the probability of an
encounter). This term depends on the global, local and community infection level. We
found that in scale-free networks the progression is slower than in random ones with the
same average connectivity. For what concerns the role of perception, we found that the
local one (information about infected neighbours) is the most effective for stopping the
spreading of the disease. However, it is also the piece of information that requires most
efforts to be gathered, and therefore it may result a high cost/efficacy ratio.

A future direction of this work is an extension of the model by inserting a balance between
the cost of information and the risk of being infected. In this regard, it should also be
taken into account what the best strategies are to avoid the spreading of epidemics in
different environments considering agents as intelligent entities capable to change or
select the best strategies dynamically in order to minimize the risk and to maximize
the economy of the system.



\section*{Acknowledgment}
We acknowledge funding from the 7th Framework
Programme of the European Union
under grant agreement  n$^\circ$ 257756 and n$^\circ$ 257906.
%
%
\bibliographystyle{plain}
\bibliography{bionetics_2012}{}

\end{document}